# Surface waves at the interface between left-handed and birefringent materials

[1,3]L. -C. Crasovan, [1]O. Takayama, [1,2]D. Artigas, [1]S. K. Johansen,

[1,3]D. Mihalache, [1,2]L. Torner

[1]*ICFO-Institut de Ciencies Fotoniques, Mediterranean Technology Park*
*08860 Castelldefels (Barcelona), Spain*
lucian.crasovan@icfo.es    http://www.icfo.es

[2] *Universitat Politecnica de Catalunya, Department of Signal Theory and Communications, 08034 Barcelona, Spain*

[3] *National Institute of Physics and Nuclear Engineering, Institute of Atomic Physics, Department of Theoretical Physics, P.O. Box MG-6, Bucharest, Romania*
http://photonics.nipne.ro

**Abstract:** We theoretically investigate the existence and properties of hybrid surface waves forming at interfaces between left-handed materials and dielectric birefringent media. The existence conditions of such waves are found to be highly relaxed in comparison to the original hybrid surface waves, discovered by Dyakonov, in configurations involving birefringent materials and right-handed media. Hybrid surface waves in left-handed materials feature remarkable properties: (i) a high degree of localization and (ii) coexistence of several guided solutions. The existence of several hybrid surface waves for the same parameter set is linked to the birefringent nature of the medium whereas the strong localization is related to the presence of the left-handed material. The hybrid surface modes appear for large areas in the parameter space.

**OCIS codes:** (240.6690) Surface waves; (260.1440) Birefringence; (160.4670) Optical materials

---

## 1. Introduction

The possibility of engineering the effective permittivity and permeability in metamaterials has attracted an increasing attention in the last decade [1-18]. Special attention has been paid to left-handed (LH) metamaterials, i.e. materials with negative permittivity and permeability as opposed to right-handed (RH) materials with positive permittivities and permeabilities, resulting in a great effort to find and engineer LH metamaterials at optical frequencies [13-18]. In such materials, highly unusual effects, e.g. negative refraction [4-8] and reversion of the Goos-Hänchen shifts [9], have been theoretically predicted and even experimentally confirmed [10,11]. The remarkable guiding properties of the negative-refractive index media have also been studied [19-24]. It has been shown that depending on the material parameters TE or TM surface waves can form at interfaces between LH metamaterials and dielectrics [19-22]. Moreover, Shadrivov *et al.* have shown by direct numerical simulations that, under appropiate conditions, a TE surface wave can be excited from an incident Gaussian beam [23]. Understanding of the physics of surface waves in general, and surface waves in metamaterials in particular, is of topical importance because of their potential applications in sensing, trapping, and imaging [25-27].

A special type of surface wave was discovered by Dyakonov in his pioneering work in 1988 [28]. Dyakonov found that surface waves can form at interfaces between two dielectrics provided that one of them is a positive birefringent medium and that the refractive index of the other isotropic medium $n_m$ lies between the two indices of the birefringent medium ($n_{ob}, n_{eb}$), i.e., $n_{ob} < n_m < n_{eb}$. Dyakonov waves feature unique properties: they are hybrid waves, i.e. they can not be decoupled into either TE or TM modes, and they only exist for a well-defined range of orientations of the optical axis with respect to the propagation direction, hereafter refered to as the *angular existence window*. However, the angular existence window is less than one degree for natural birefringent materials. This makes the experimental excitation a challenge, and these surface waves have not yet been experimentally observed. Several strategies have been suggested to increase the angular existence domain, such as the use of a film some nanometers thick in between the two dielectrics [29-32], the use of photonic metamaterials featuring huge effective birefringences and tuning capabilities [33], or the use of isotropic magnetic media [34]. Dyakonov surface waves also form in biaxial birefringent crystals [35].

Motivated by these findings, we explore the existence of hybrid surface waves, in some sense similar to Dyakonov waves, at interfaces between LH and RH media. Thus, we investigate the existence and features of the surface waves forming at interfaces between RH birefringent dielectrics and LH materials. We show that, in contrast to Dyakonov surface waves existing in RH media, for some material parameters it is possible to have hybrid surfaces waves without any restriction on the orientation angle of the birefringent material. In addition, we show that for given conditions, diverse surface waves can coexist, and that some of these solutions can be highly localized, similarly to plasmons.

The paper is organized as follows. In Chapter 2, we put forward the eigenvalue equation for the surface waves at interfaces between LH and uniaxial media and introduce the parameters of the problem. Chapter 3 presents the results on the existence and properties of the hybrid surface waves, together with a brief comparison with previous studies. In Chapter 4 we present the concluding remarks and discuss the experimental implications of the present study.

## 2. Eigenvalue equation for surface waves at birefringent-metamaterial interfaces

We consider a planar interface separating two semi-infinite media: an uniaxial RH medium and a LH metamaterial. We assume that the optical axis of the birefringent material lies in the $(x,y)$-interface-plane, making an angle $\theta$ with the propagation direction $x$. The $z$ axis is perpendicular to the interface. Throughout the paper we assume harmonic electromagnetic plane waves with the time dependence of all the electric and magnetic field components being $\exp(-i\omega t)$. We let $\varepsilon_{ob}, \varepsilon_{eb}, \varepsilon_m$ and $\mu_b, \mu_m$ be the relative permittivities and permeabilities corresponding to the ordinary/extraordinary waves in the uniaxial medium and to the metamaterial, respectively. We will assume equal permeabilities for the ordinary and the extraordinary waves. The corresponding refractive indices are $n_{ob,eb} = \sqrt{\varepsilon_{ob,eb}}\sqrt{\mu_b}$ and $n_m = \sqrt{\varepsilon_m}\sqrt{\mu_m}$, respectively. With this definition one gets a negative value for the metamaterial refractive index if the permittivity and permeability are simultaneously negative.

The wave equation for the electric field $\boldsymbol{E} = (E_x, E_y, E_z)^T$ in a uniaxial medium writes

$$\nabla^2 \boldsymbol{E} + k_0^2 \hat{\varepsilon} \boldsymbol{E} = \nabla(\nabla \cdot \boldsymbol{E}) , \qquad (1)$$

where $k_0$ is the vacuum wavenumber and $\hat{\varepsilon}$ the permittivity tensor that, for the above mentioned configuration, has only five nonvanishing elements: $\varepsilon_{xx} = n_{ob}^2 \sin^2\theta + n_{eb}^2 \cos^2\theta$, $\varepsilon_{yy} = n_{ob}^2 \cos^2\theta + n_{eb}^2 \sin^2\theta$, $\varepsilon_{xy} = \varepsilon_{yx} = (n_{eb}^2 - n_{ob}^2)\sin\theta\cos\theta$ and $\varepsilon_{zz} = n_{ob}^2$. The electromagnetic fields are proportional to $\exp(-ik_0 N x)$, where $N$ is the effective refractive index of the surface wave and is obtained as the eigenvalue solution of the boundary condition problem. After expressing the electric and magnetic fields as linear combinations of TE and TM modes in the metamaterial and as ordinary and extraordinary modes in the birefringent medium, respectively, we impose continuity of the tangential components of the electric and magnetic fields $(E_x, E_y, H_x, H_y)$ at the interface and end up with the eigenvalue equation for the surface modes (see [32] for further details). Assuming evanescent waves in the $z$ direction, the eigenvalue equation reads:

$$n_{ob}^2 A_e B_o \sin^2\theta - \gamma_{ob}^2 A_o B_e \cos^2\theta = 0 , \qquad (2)$$

where we have defined:

$$A_{o,e} = \frac{\gamma_{ob,eb}}{\mu_b} + \frac{\gamma_m}{\mu_m}, B_o = \frac{n_m^2 \gamma_{ob}}{\mu_m} + \frac{n_{ob}^2 \gamma_m}{\mu_b}, B_e = \frac{n_m^2 \gamma_{ob}}{\mu_m} + \frac{n_{ob}^2 \gamma_m \gamma_{eb}}{\mu_b \gamma_{ob}} ,$$

$$\gamma_m = (N^2 - n_m^2)^{1/2}, \gamma_{ob} = (N^2 - n_{ob}^2)^{1/2}, \gamma_{eb} = \frac{n_{eb}}{n_{eb}(\theta)}(N^2 - n_{eb}^2(\theta))^{1/2} , \qquad (3)$$

$$n_{eb}(\theta) = \frac{n_{eb} n_{ob}}{(n_{ob}^2 \sin^2\theta + n_{eb}^2 \cos^2\theta)^{1/2}} .$$

The above eigenvalue equation is general and holds for any non-lossy surface wave forming at a two-media interface independently of the refractive index sign. For interfaces between LH and RH media the solutions will differ considerably from those existing at interfaces between two RH materials. In contrast, interfaces between two RH or two LH materials result in the same eigenvalue equation, and therefore behave similar properties.

We will restrict ourselves in what follows to the case of metamaterials having both negative permittivity and negative permeability. In principle, this regime can be reached when the metamaterial is operated at a frequency beyond the plasma and magnetic resonance frequencies. In addition, the operating frequency must be below the magnetic plasma frequency to avoid losses. If one fixes the absolute values of the refractive indices $n_{ob,eb,m}$ the solutions of Eq. (2) depend only on the permeability ratio $r = \mu_m / \mu_b$ and on the orientation angle $\theta$. Moreover, in the limit cases $\theta = 0°$ and $\theta = 90°$ we see that Eq. (2) reduces to $A_e B_o = 0$ and $A_o B_e = 0$, respectively, corresponding to decoupling of the hybrid modes into TE and TM modes.

### 3. Surface wave solutions

We will now present the results obtained on the existence and properties of the hybrid surface wave solutions. In order to find the solutions for each set of parameters $(n_{ob,eb,m}, r, \theta)$, we have numerically solved Eq. (2). For illustrative purposes we have throughout the paper fixed the refractive indices of the positive birefringent medium to $n_{ob} = 1.52$ and $n_{eb} = 1.725$, corresponding to the values of an E7 liquid crystal at 632.8 nm. Qualitatively similar results were obtained for other values of the refractive indices of the birefringent medium. We note that hybrid surface waves were also found to form at interfaces between LH materials and uniaxial media with negative birefringence, i.e. $n_{eb} < n_{ob}$, as opposed to the Dyakonov surface waves forming at interfaces between RH isotropic and positive birefringent materials. A detailed investigation of this situation is beyond the scope of this paper and will be reported elsewhere.

First, we analyze the existence conditions of the surface waves. The results appear in Figs. 1 and 2 as regions where solutions are allowed or forbidden. The separation curves correspond to the cut-off values with $N = |n_m|$, $N = n_{eb}(\theta)$ and $N \to \infty$. Such cut-off conditions also hold for the conventional Dyakonov surface waves forming at interfaces between positive birefringent and isotropic media and for ultranarrow sandwich structures that host hybrid guided modes below the usual cut-off. From the eigenvalue equation one can directly write down the equations these cut-off values have to fulfill. For the $N = |n_m|$ cut-off one obtains:

$$\sin^2\theta \; n_{ob}^2 \; \gamma_{eb} - \cos^2\theta \; \gamma_{ob}^3 = 0 \quad , \tag{4}$$

whereas at $N = n_{eb}(\theta)$ one gets:

$$\sin^2\theta \; n_{ob}^2 \gamma_m (n_m^2 \gamma_{ob} + r\, n_{ob}^2 \gamma_m) - \cos^2\theta \; n_m^2 \gamma_{ob}^3 (\gamma_m + r\, \gamma_{ob}) = 0 \quad . \tag{5}$$

Finally, when $N \to \infty$ we can directly express $r$ as:

$$r = -1, \quad r = -\frac{n_m^2 n_{eb}(\theta)}{n_{ob}^2 n_{eb}} \quad . \tag{6}$$

Solving Eqs. (4) and (5) for $\theta$ (or $r$) by fixing $r$ (or $\theta$) one gets the border $\theta(n_m)$ (or $r(n_m)$). Note that Eq. (4) does not contain $r$, therefore only a $\theta(n_m)$ border can be found. To guide the eye we have plotted these demarcation curves as colored lines in all the panels of Figs. 1 and 2.

Fig. 1 corresponds to representations in the $(|r|, |n_m|)$-parameter plane for different orientations of the optical axis. In addition, we have included in Fig. 1(a) the equivalent results for an interface between a LH and an isotropic medium, situation analyzed in Refs. [19-22]. This figure clearly shows the existence of surface waves when at least one of the two media is a LH metamaterial. This is clearly in contrast with the case where both media are RH materials, where at least one of the media must be berefringent in order to obtain Dyakonov waves. When comparing Fig. 1(a) with figures corresponding to the limit cases $\theta = 0°$ (Fig. 1(b)) and $\theta = 90°$ (Fig. 1(f)), the demarcation curves split the parameter plane into regions where only TE-polarized, only TM-polarized, or both types of modes coexist, together with the white regions at which there are no solutions. Note, that in the case of the interface between a LH and an isotropic medium TE- and TM-polarized modes can not coexist. On the contrary, at interfaces between LH and birefringent materials they can coexist at both $\theta = 0°$ and $\theta = 90°$. The coexisting regions of the parameter plane are shown by the dark-gray regions in Fig. 1(b)-(f) and appears at $n_{ob} < |n_m| < (n_{ob} n_{eb})^{1/2}$ for $\theta = 0°$ and $n_{ob} < |n_m| < n_{eb}$ for $\theta = 90°$. These new regions where both TE and TM surface waves coexist are consequently linked to the birefringent nature of the medium. Coexistence of two surface modes constitutes a rare, though not unique, situation in physics, e.g. coexistence has also been predicted in configurations using gyrotropic-LH material interfaces [24].

When changing the orientations of the optical axis, the surface waves are no longer purely TE- or TM-polarized, but hybrid modes. Moving away from $\theta = 0°$ the cut-off curve corresponding to $N = n_{eb}(\theta)$ splits into two lines (see the arrows in Fig. 1(b) and the blue lines in Figs. 1(c)-(f)) that move away from each other giving birth to a region in the parameter plane where, for moderate small orientation angles ($\theta \leq 30°$), single solution surface waves exist only if the absolute value of the relative permeability $|r|$ is slightly smaller than unity. As the angle approaches $\theta = 90°$, new regions where two different hybrid solutions coexist appear at $|r| > 1$, while coexisting solutions at $|r| < 1$ progressively disappear. Overall, we observe that the coexisting surface waves basically appear inside the existence domain $n_{ob} < |n_m| < n_{eb}$ where Dyakonov waves in RH materials exist. Only, small areas outside this existence domain at $|n_m| < n_{ob}$ with $|r| < 1$ break this rule (see Figs. 1(c)-(e)).

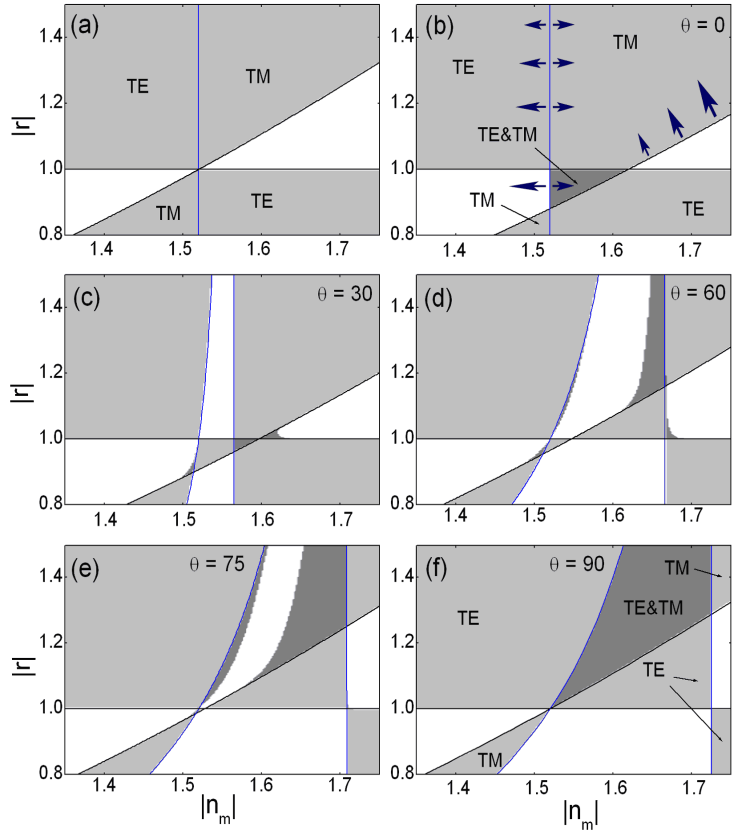

Fig.1 Existence domains in the $(|r|, |n_m|)$ plane for (a) LH metamaterial-isotropic interface and for LH metamaterial-birefringent interfaces at (b) $\theta = 0°$, (c) $\theta = 30°$, (d) $\theta = 60°$, (e) $\theta = 75°$, and (f) $\theta = 90°$. Blue lines: $N = n_{eb}(\theta)$; black lines: $N \to \infty$. White regions: no solutions. Light gray, dark gray and black regions correspond to single mode, 2-mode, and 3-mode regions, respectively.

Figure 2 shows the existence domains in the $(\theta, |n_m|)$ parameter-plane for fixed values of the relative permeability $r$. In the most part of the existence domain one surface wave solution exists (see the light-gray regions in all the panels of Fig. 2). As $|r|$ increases, the cut-off line moves towards higher values of $|n_m|$ and more surface waves start to appear, forming isolated regions where two (dark-gray regions) or rarely three (black regions in Fig. 2(e)) solutions coexist. Note the large angular domains (tens of degrees) and range of $|n_m|$, especially at $|r|>1$, for which these coexisting surface waves appear. This implies that such waves should be easily excited and observed in configurations with metamaterials.

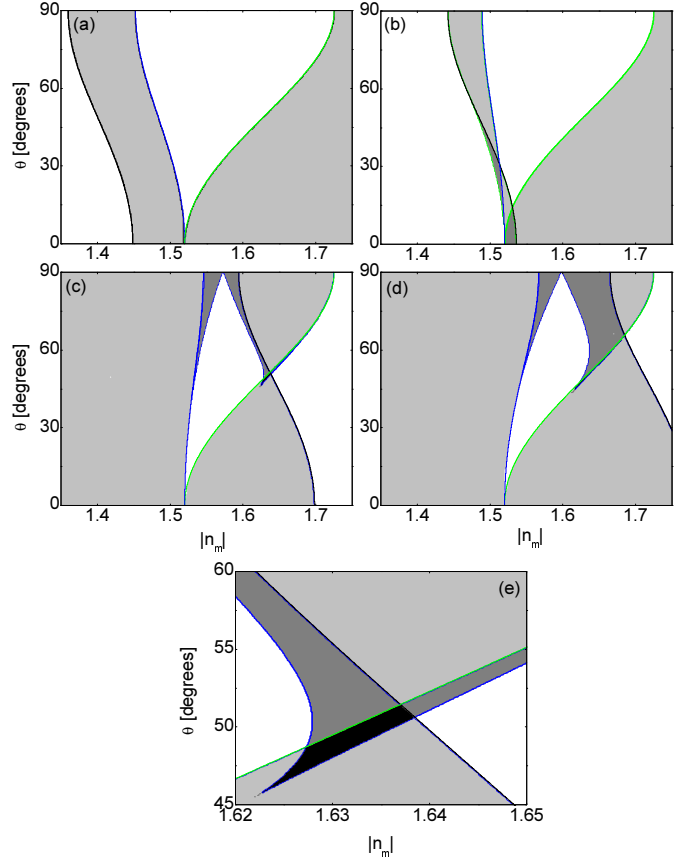

Fig.2 Existence domains in the $(\theta, |n_m|)$ plane for several values of the relative permeability. (a) $|r|=0.8$, (b) $|r|=0.9$, (c) $|r|=1.1$, and (d) $|r|=1.2$. (e) Zoom of panel (c). Blue lines: $N = n_{eb}(\theta)$, green lines: $N = |n_m|$ and black lines: $N \to \infty$. Gray color scale has the same meaning as in Fig. 1.

In order to illustrate the dependence of the eigenvalue on the absolute value of the refractive index in the metamaterial, $N(|n_m|)$, some modal diagrams are shown in Fig. 3 for two representative values of the relative permeability. Fig. 3(a) displays modal regions where two and three modes coexist. Typically, the branches terminate on either the $N = n_{eb}(\theta)$, the $N = |n_m|$, or the $N \to \infty$ cut-off curve. Moreover, in a region where two or three solutions coexist, a new cut-off line appears at points where two of the solutions meet or merge. When plotting $N(|n_m|)$, these points appear as turning points where the upper branch of the curves turn into the lower parts. Turning points are a consequence of the intrinsic hybrid nature of the surface wave that form at $\theta \neq 0°$. However, at $\theta = 90°$ these characteristic points dissapear, resulting in the two TE and TM solutions crossing each other (see Fig. 3(b)). We have also followed the evolution of the eigenvalues of the surface waves along a constant $|n_m|$ in Fig. 2, thus plotting $N(\theta)$. These dependencies are shown in Figs. 3(c) and 3(d) for two values of the LH metamaterial refractive index at relative permeability of $r = -1.1$ and $r = -1.2$. They complement Figs. 3(a) and 3(b) showing in more detail the existence range for each coexisting solution in terms of the orientation angle.

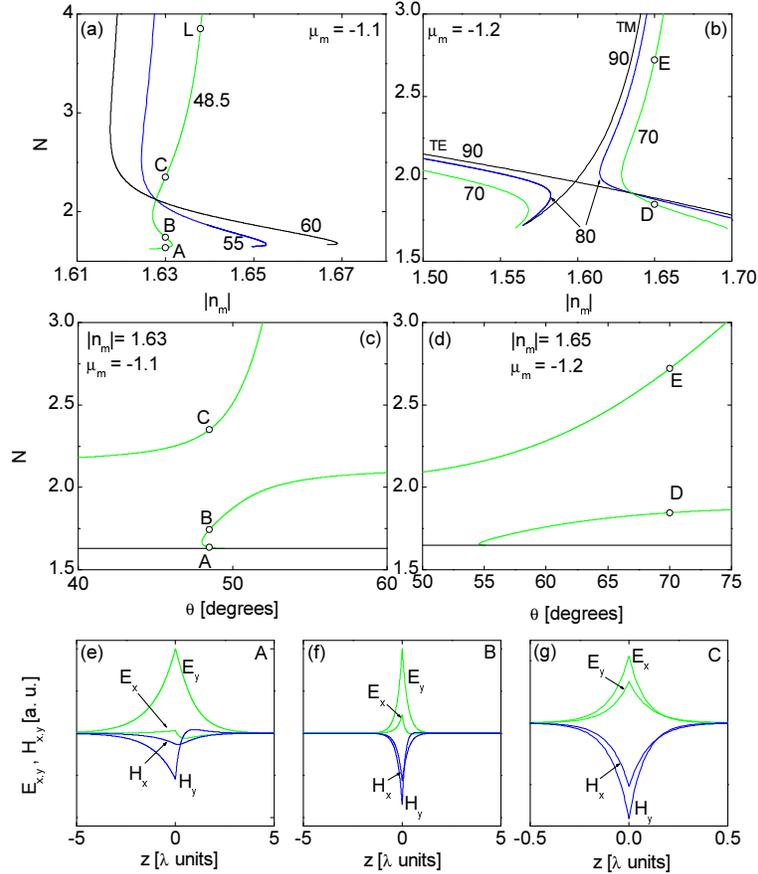

Fig.3 Mode eigenvalue versus metamaterial refractive index ((a) and (b)) or orientation angle ((c) and (d)). In (a) and (c) $|r|=1.1$, whereas in (b) and (d) $|r|=1.2$. The green, blue, and black lines correspond to $\theta = 48.5°$, $\theta = 55°$, and $\theta = 60°$ in panel (a) and to $\theta = 70°$, $\theta = 80°$, and $\theta = 90°$ in panel (b), respectively. Labels close to lines indicate orientation angle in degrees. Black lines in panels (c) and (d): $N = |n_m|$ cut-offs. (e)-(g) Field components of the coexisting solutions labeled A through C in panel (a). Here green and blue lines stand for the electric and magnetic field components, respectively. Note the smaller scale in panel (g) due to the strong confinement of the field.

The three solutions labeled A, B and C, shown in Fig. 3(a) and 3(c) coexist for $r = -1.1$ and $n_m = -1.63$. These solutions feature significant different effective refractive indices, field distributions, and degrees of localization at the interface. We show their corresponding profiles in Figs. 3 (e)-(g). Among them, solution C features the strongest localization, i.e. it is confined within half of the wavelength. This solution corresponds to a branch that ends on the cut-off line $N \to \infty$. Solution A, being close to the $N = |n_m|$ cut-off, is the most delocalized, spreading over an interval of ten wavelengths. Finally, solution B, which is further away from the cut-off, has a degree of localization on the order of the wavelength. To give a measure of the hybrid nature of the solutions, we plot in Fig. 4 the dependence of the TE/TM fraction in terms of the orientation angle for different material parameters. Large values of the ratio $\eta = |E_y(z=0)/E_x(z=0)|$ indicate a TE-dominant wave, whereas small values ($\eta \ll 1$) indicate a TM dominance. We can see the different TE or TM dominant nature of the three and two coexisting solutions in Fig. 4(a) and 4(b), respectively. This shows that all regimes, i.e. TE-dominant, TM-dominant or TE/TM balanced waves, can be met by tuning the material parameters and the orientation angle of the optical axis.

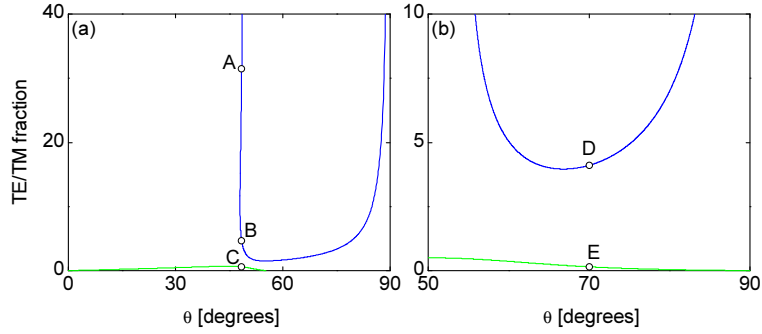

Fig.4 TE/TM fraction ( $|E_y(z=0)/E_x(z=0)|$ ) versus orientation degree $\theta$ for representative values of metamaterial refractive index. (a) corresponds to $|n_m|=1.63$ and $|r|=1.1$ ; (b) to $|n_m|=1.65$ and $|r|=1.2$ . Labels A through E correspond to points A-E in Figs. 3(a)-3(d).

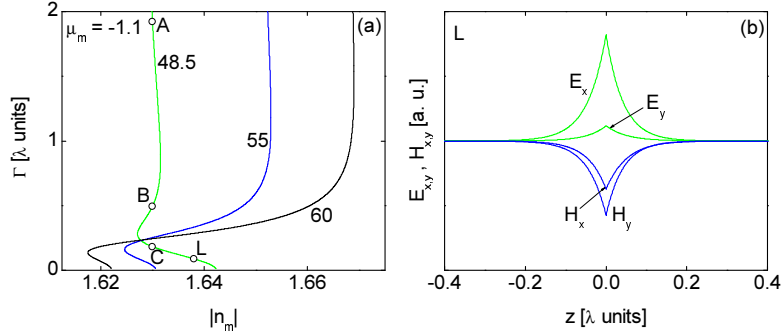

Fig.5 (a) Degree of localization $\Gamma$ versus $|n_m|$. Parameters corresponds to Fig. 3(a). In (b) the field components corresponding to the highly localized solution labeled L in Fig. 3(a) are shown. Green and blue lines correspond to the electric and magnetic field components, respectively.

One of the properties mentioned above, the high degree of localization at the surface, is an important feature that makes these surface waves potentially interesting for future applications. Strong localization appears at both isotropic-LH and birefringent-LH interfaces, a feature that has not been properly appreciated to date. Conceptually, this strong localization is similar to the one found with surface plasmons [27]. In general, any solution corresponding to $N \gg |n_{m,eb}|$ in Figs. 1 and 2 will correspond to surface waves confined well below the wavelength limit, since in these situations, the field decay constants increase approximately linearly with $N$. In order to illustrate this strong localization we show in Fig. 5(a) the degree of localization, defined as the surface wave width at $1/e$ from the maximum amplitude, corresponding to Fig. 3(a). Here we can see that the width decreases for solutions in the $N \gg |n_{m,eb}|$ branch. An example of a highly localized surface wave, less than one-tenth of the wavelength, corresponding to point L in Figs. 3(a), is shown in Fig. 5(b). Notice the TM dominant character of this solution. When $N \gg |n_{m,eb}|$, the surface waves become more and more TM dominant, showing its plasmon-like nature.

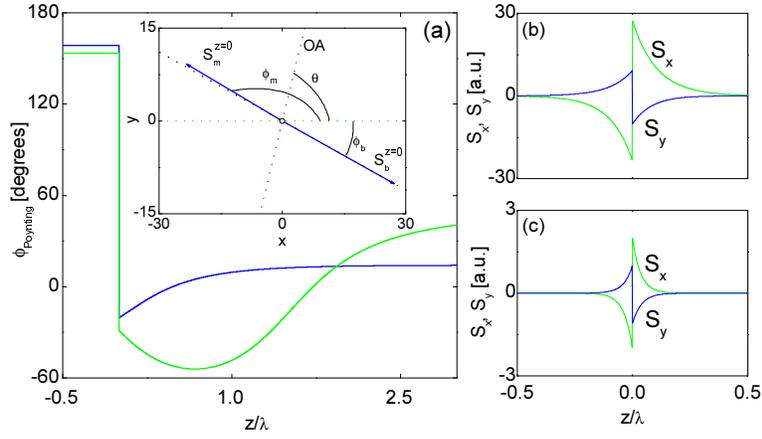

Fig.6 (a) Walk-off angle of the Poynting vector (the angle the Poynting vector forms with the $x$-axis) across the interface for the surface waves corresponding to points D (blue line) and E (green line) in Figs. 3 and 4. Inset shows the Poynting vectors for solution D in the two neighboring media in close proximity of the interface. Red, green, and black dotted lines in the inset stand for the optical axis, the $x$-axis, and the direction of the Poynting vector in the birefringent media close to the interface, respectively. The transverse components of the Poynting vectors versus $z$ for solution D and E are shown in (b) and (c) respectively.

As the surface waves are exponentially decaying away from the interface with no real component of the wave number along the $z$-direction, energy propagates only in the $(x,y)$-plane. It has been shown that at interfaces between isotropic LH and RH media, the Poynting vector of the surface waves changes sign [19-22]. As our geometry involves a birefringent medium, one would expect to have a non-zero walk-off angle between the direction of the propagation of the energy and the wave vector (assumed to be along $x$-axis in our case) and, additionally, sign jumps of the Poynting vector components across the interface due to the presence of the negative refractive index metamaterial. We have calculated the Poynting vector components and the angle formed by the Poynting vector with the $x$-axis for two typical coexisting surface waves, and indeed we find that a walk-off angle exists in both neighboring media. Moreover, in the metamaterial the walk-off angle is constant at all distances to the interface, whereas in the birefringent medium the Poynting vector rotates in a different fashion for the two considered solutions when moving away from the interface. The difference in the Poynting vector rotation is due to the different decay constants and amplitudes featured by the transverse field components (see Fig. 6) in the two considered solutions. Finally, another characteristic of these hybrid surface waves is that the Poynting vector in the close vicinity of the interface between the two media is not completely anti-parallel as it is for the LH-isotropic interface, i.e. a small deviation of a few degrees is noticed.

## 7. Conclusion

In conclusion we have uncovered, for the first time to our knowledge, the existence of surface waves at the interface between LH metamaterials and birefringent media. We have found that, apart from the solutions appearing at interfaces between LH and isotropic materials, a new family of different coexisting solutions exist. Such new solutions are linked to the birefringent nature of the dielectric media and result in hybrid surface waves. In this configuration, we discovered that different hybrid surface waves solutions can usually coexist in groups of two or three for the same governing parameters. Most of these coexisting solutions appear for LH refractive indices that fulfill the condition $n_{ob} < |n_m| < n_{eb}$. However, their angular existence window is significantly larger than the corresponding window for Dyakonov waves forming in birefringent RH media. Moreover, under certain conditions, the surface waves at LH interfaces were found to be highly localized, i.e., featuring a localization degree similar to that of plasmons. Finally, the Poynting vectors in the two neighboring media are not completely anti-parallel and rotate in the birefringent medium as a function of the distance to the interface.

**Acknowledgements**

This work is supported in part by the Ministry of Science of the Government of Spain through the *Juan de la Cierva* programme (L.-C.C.).